\documentclass[showpacs,amsmath,amssymb]{revtex4}
       
\usepackage{bbold}
\usepackage{epsfig}
\usepackage{graphpap}

\newcommand{\ru}[3]{\rule[#1mm]{#2mm}{#3mm}}

\newcommand{\Feff}{F^{\mbox{\footnotesize eff}}}
\newcommand{\cFeff}{{\cal F}^{\mbox{\footnotesize eff}}}

\newcommand{\eps}{\varepsilon}
\newcommand{\lam}{\mbox{\bbfamily \char 21}}

\begin{document}
\setlength{\unitlength}{1mm}

\title{Synergetic System Analysis for the Delay-Induced Hopf Bifurcation\\
in the Wright Equation}

\author{Michael Schanz}

\address{Institute of Parallel and Distributed
High-Performance Systems (IPVR), University of Stuttgart, \\
Breitwiesenstra{\ss}e 20-22, D-70565 Stuttgart, Germany\\
E-mail: {\tt michael.schanz@informatik.uni-stuttgart.de} }

\author{Axel Pelster}

\address{Institute of Theoretical Physics, Free University of Berlin, Arnimallee 14, D--14195 Berlin, Germany\\
E-mail: {\tt pelster@physik.fu-berlin.de}\\$\mbox{}$ }

\date{\today}

\begin{abstract}
We apply the synergetic elimination procedure for the
stable modes in nonlinear delay systems close to a dynamical instability and
derive the normal form for the delay-induced Hopf bifurcation
in the Wright equation. The resulting periodic orbit is confirmed by numerical simulations.
\end{abstract}
\maketitle
\section{Introduction}
Within the last decades synergetics has provided powerful concepts
and methods to describe selforganization processes in various branches
of science \cite{Haken1,Haken2,Haken3,Haken4,Haken5}. The spontaneous
formation of spatial, temporal or functional patterns in complex systems
has been successfully investigated by working out general principles and
by mapping them onto universal mathematical structures. The important
result is due to the fact that in the vicinity of 
a dynamical instability
the high-dimensional set of nonlinear evolution equations modelling a 
complex system on a microscopic or a mesoscopic scale can be 
approximately reduced to a low-dimensional set of order parameter equations
describing the evolving pattern formation on a macroscopic scale. To
obtain such a simplified, reduced description of selforganization processes,
the synergetic system analysis proceeds as follows. A linearization
of the evolution equations around a stationary solution shows that a 
dynamical instability is always accompanied by a time-scale hierarchy
between numerous fast modes $s$ and few slow modes $u$. A rigorous treatment
of the full nonlinear evolution equations in the vicinity of the dynamical
instability leads to a characteristic interdependence between both hierarchy
levels which may be illustrated by a circular causality chain.
On the one hand the slaving principle of synergetics states that the numerous
fast modes $s$ quasi-instantaneously take values which are prescribed by the
few slow modes $u$ according to 
$s ( t ) = h ( u ( t ) )$ with the center manifold $h ( u )$.
On the other hand an adiabatic elimination of the fast enslaved modes $s$
yields equations for the slow order parameters $u$ which depend, in general,
on the center manifold $h ( u )$ due to the nonlinear feedback.\\

In its original formulation, the synergetic system analysis was developed
for complex systems which can be modeled by ordinary and partial differential
equations as well as their stochastic generalizations. Some time ago,
the general concepts and methods of synergetics have been extended to delay
differential equations to deal with dynamical instabilities which are induced
by the finite propagation time of signals in feedback loops \cite{Wolfgang}.
Taking into account the infinite-dimensional character of a delay system
\cite{Hale,Krasovskii}, the adiabatic elimination of the stable modes leads to a
low-dimensional set of order parameter equations which turn out to be of
the form of ordinary differential equations, i.e. they no longer contain
memory effects. The predictions of the synergetic system analysis have been
quantitatively tested by investigating the delay-induced Hopf bifurcation
of the electronic system of a first-order phase-locked loop (PLL)
\cite{Wolfgang}. The periodic orbit which results from the corresponding
order parameter equation near the bifurcation point has been confirmed
by both a multiple scaling procedure and numerical simulations 
\cite{Michael1,Michael2}. Although this application exemplarily proves the
order parameter concept for delay systems, it does not allow to draw
conclusions about the slaving principle. As the lowest nonlinear term
in the scalar delay differential equation of the PLL is a cubic one, the
center manifold does not influence the order parameter equation
of the Hopf bifurcation in the lowest order. In order to check for delay systems both ingredients of the
circular causality chain, i.e. the order 
parameter concept and the slaving principle, 
it is thus indispensible to 
study a scalar delay differential equation with a quadratic nonlinearity. \\

A candidate for such a study is provided by
\begin{eqnarray}
\label{verhulst}
\frac{d}{dt} z ( t ) = R \, \left[ z ( t ) - z ( t - \tau )^2 \right] \, .
\end{eqnarray}
With vanishing time delay $\tau$ it represents a system which is named after the Belgian mathematician P.F. Verhulst 
from the 19th century \cite{verhulst1}. It is used as a simplified model for the population dynamics of a species in an
environment with limited food supply \cite{verhulst2}. The synergetic system analysis for the Verhulst system with time
delay (\ref{verhulst}) has been already performed in Ref.~\cite{Michael1}. 
There also the well-known equation of Wright \cite{Wright}
\begin{eqnarray}
\label{g-wr1-syso}
\frac{d}{dt} z(t) &=& - R z(t-\tau) \left[ 1 + z(t) \right]
\end{eqnarray}
has been treated, 
where $R$ denotes a system parameter and $\tau$ a delay time.
This delay differential equation is mentioned by Wright \cite{Wright} as arising
in the application of probability methods to the theory of asymptotic prime number
density. Cunningham \cite{Cunningham} depicts it as a ``growth equation'' representing
a mathematical description of a fluctuating population of organisms under certain
environmental conditions. In addition, it may describe the operation of a control
system working with potentially explosive chemical reactions, and quite similar
equations arise in economic studies of business cycles. 
Performing an appropriate scaling of time
\begin{eqnarray}
t = \tau \, t' \, , \quad z' ( t' ) = z ( \tau \, t' ) 
\end{eqnarray}
converts the Wright equation (\ref{g-wr1-syso}) to its standard form with the control parameter 
\begin{eqnarray}
R' = \tau \, R  \, .
\end{eqnarray}
Thus varying the delay time $\tau$ corresponds to changing the control parameter $R'$. By omitting the prime ' for the
respective quantities, the standard form of the Wright equation reads
\begin{eqnarray}
\label{g-wr1-sys}
\frac{d}{dt} z(t) &=& - R z(t-1) \left[ 1 + z(t) \right] \, .
\end{eqnarray}
In this paper we restrict ourselves to analyse this standard form of the Wright equation.\\

The Wright equation (\ref{g-wr1-sys}) shows a delay-induced instability, namely a
Poincar\'e-Andronov-Hopf bifurcation at the critical value
\begin{eqnarray}
\label{crit}
R_c = \frac{\pi}{2}
\end{eqnarray}
of the control parameter $R$.
In Ref. \cite{Marsden} it is shown that the oscillatory solution in the vicinity of this
instability, i.e. the emerging limit cycle,
can be calculated approximately using the method of {\em averaging}.
This approximation reads in the lowest order
\begin{eqnarray}
\label{g-wr1-vgl}
z(t) &=& A \, \sqrt{R - \frac{\pi}{2}} \, \cos \left( \frac{\pi}{2} t \right)
+ {\cal O} \left( R - \frac{\pi}{2} \right) \, ,
\end{eqnarray}
where the amplitude $A$ has the value
\begin{eqnarray}
\label{g-wr1-vglb}
A = \sqrt{\frac{40}{3 \pi - 2}} \, .
\end{eqnarray}
In Section \ref{la} we start with a linear stability analysis of the Wright equation (\ref{g-wr1-sys}) which
confirms, of course, the delay-induced Poincar\'e-Andronov-Hopf bifurcation
when the control parameter $R$ approaches the critical value (\ref{crit}).
Near this instability we perform a nonlinear synergetic treatment in Section III
and study in detail how the center manifold influences
the order parameter equation. In Section IV the resulting order parameter equation is transformed to the normal form
of a Hopf bifurcation, where the emerging periodic orbit is determined one order higher
than the lowest-order result (\ref{g-wr1-vgl}) and (\ref{g-wr1-vglb}). The numerical investigations of Section V
confirm the emerging periodic orbit, furthermore we discuss 
the global bifurcation scenario of the Wright equation (\ref{g-wr1-sys}).

\section{Linear Stability Analysis}
\label{la}

The solution of the delay differential equation (\ref{g-wr1-sys}) for times $t\ge0$ depends on the initial values
of the function $z(t)$ in the entire interval $[-1,0]$. In order to properly define such an initial value problem,
Hale \cite{Hale} and Krasovskii \cite{Krasovskii} proposed to transform
the equation of motion (\ref{g-wr1-sys}) for a function $z(t)$ in the usual state space
$\Gamma$ to the extended state space
${\cal C}$ of continuous complex valued functions $z_t(\Theta)$ which are defined
on the interval $[-1,0]$:
\begin{eqnarray}
\label{g-wr1-syse}
\frac{d}{dt} z_t(\Theta) &=&
\left( {\cal G} \, z_t \right)(\Theta) = \left\{
\begin{array}{l@{\quad,\quad}c}
{\displaystyle \frac{d}{d \Theta}} z_t(\Theta) & -1 \le \Theta < 0 \\[3mm]
{\cal F} [z_t] & \Theta = 0
\end{array} \right. \, .
\end{eqnarray}
Following the notation of Ref. \cite{Wolfgang}, we introduced not only the new function
$z_t \in {\cal C}$, which is connected to the
original function $z(t) \in \Gamma$ through 
\begin{eqnarray}
\label{map}
z_t(\Theta) = z(t+\Theta)\,, \quad -1 \le \Theta \le 0\,,
\end{eqnarray}
but also the nonlinear functional
\begin{eqnarray}
\label{g-wr1-sysef}
{\cal F} \left[ z_t\right] &=&
\sum_{k=1}^{2} \int_{-1}^0 d \Theta_1 \cdots \int_{-1}^0 d \Theta_k
\, \omega^{(k)}(\Theta_1,\ldots,\Theta_k) \, \prod_{l=1}^k z_t(\Theta_l)
\end{eqnarray}
with the two scalar densities
\begin{eqnarray}
\label{g-wr1-sysed1}
\omega^{(1)}(\Theta_1) &=& -R \mbox{\boldmath $\delta$}(\Theta_1+1) \, ,\\
\label{g-wr1-sysed2}
\omega^{(2)}(\Theta_1,\Theta_2) &=& -R \mbox{\boldmath $\delta$}(\Theta_1+1)
\mbox{\boldmath $\delta$}(\Theta_2) \, .
\end{eqnarray}
The stationary states of this system
\begin{eqnarray}
\label{g-wr1-stat-1}
z_{\rm stat}^{\rm I} = 0 \, , \quad\quad z_{\rm stat}^{\rm II} = -1 
\end{eqnarray}
are candidates for the reference state from which we start our further
investigations. For the main body of the article we focus our attention on the stationary state
$z_{\rm stat}^{\rm I}$ and choose it as the reference state.
The other stationary state $z_{\rm stat}^{\rm II}$ will be discussed in 
Section V together with the global bifurcation scenario of the Wright equation (\ref{g-wr1-sys}).\\

\begin{figure}
\setlength{\unitlength}{1mm}
\begin{center}
\begin{picture}(120,80)
%\graphpaper[5](0,0)(120,80)
%\put(-10,66){\epsfig{figure=lin_pll.eps,width=100\unitlength,angle=-90}}
\put(-15,90){\epsfig{figure=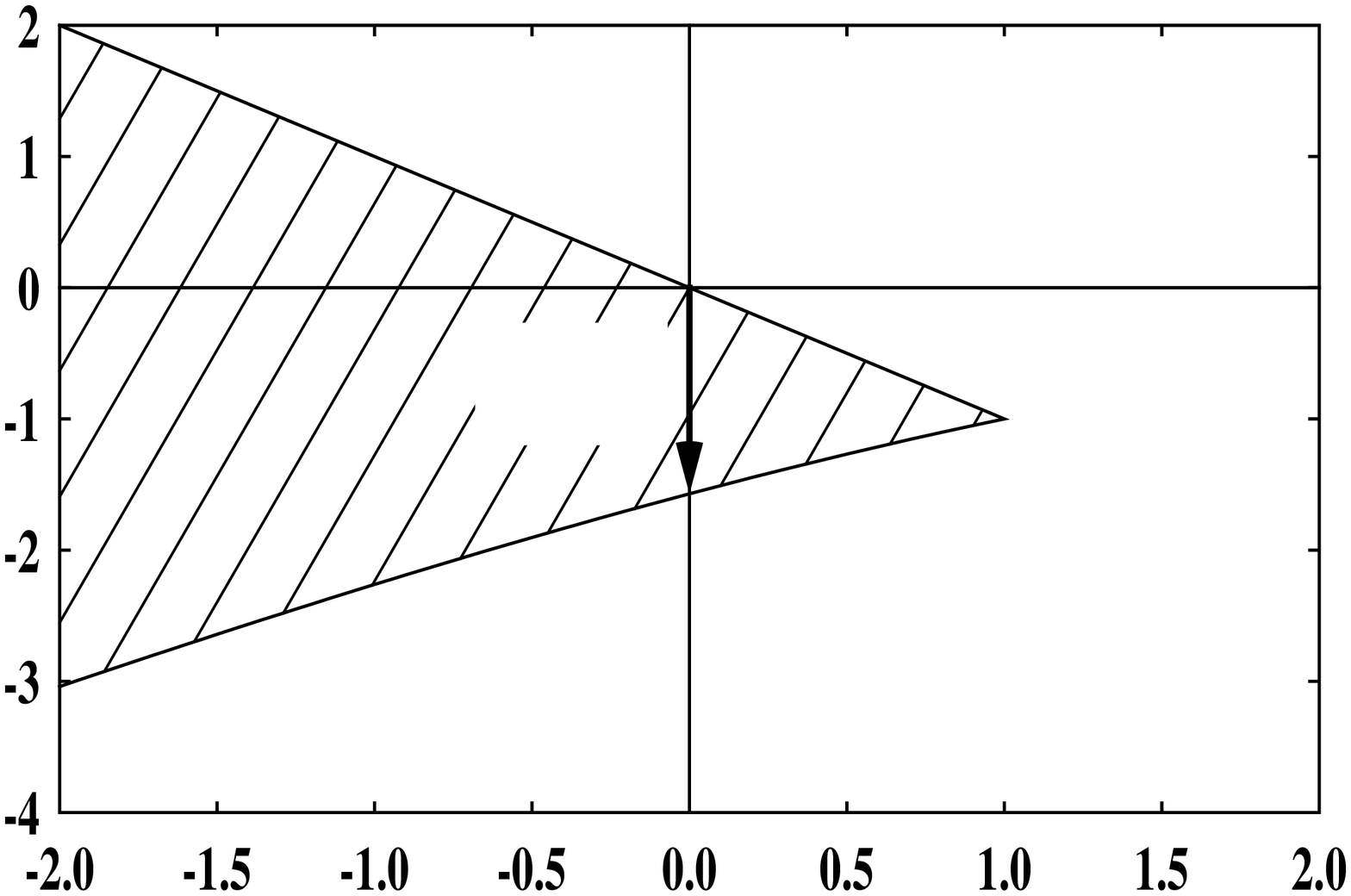,width=100\unitlength,angle=-90}}
\put(59,-2){$p$}
\put(0,40){$q$}
\put(48,42){$z_{\rm stat}^{\rm I}$}
\end{picture}
$\mbox{}$\\[3mm]
\caption{\label{b-hayes}Within the shaded region both conditions of the Hayes theorem 
(a) $p < 1$ and (b) $p \, < \, -q \, < \, \sqrt{a_1^2 + p^2}$ with $a_1 = p \tan (a_1)$ 
lying in the interval $[0,\pi)$ are fulfilled. Linearizing the Wright equation (\ref{g-wr1-sys}) around the stationary
state $z_{\rm stat}^{\rm I}=0$ we have $p = 0$ and $q = -R$. When the control parameter $R$ is increased from $0$ to
$\pi/2$, we move in the $q,p$-plane along the arrow within the shaded stability region.
Thus all solutions of the characteristic equation (\ref{g-wr1-char}) have negative
real part. The boundary of the shaded stability region is reached at the instability $R_c=\pi/2$.}
\end{center}
\end{figure}

Then we linearize the system (\ref{g-wr1-syse}) with respect to the
stationary state $z_{\rm stat}^{\rm I} = 0$ by using the decomposition
\begin{eqnarray}
z_t(\Theta) = z_{\rm stat}^{\rm I} + \zeta_t(\Theta) \, , \quad -1 \le \Theta \le 0 \, .
\end{eqnarray}
This leads to the following linearized equation of motion for the deviation $\zeta_t(\Theta)$
from the stationary state $z_{\rm stat}^{\rm I}=0$:
\begin{eqnarray}
\label{g-wr1-sysel}
\frac{d}{dt} \zeta_t(\Theta) &=&
\left( {\cal G}_L \, \zeta_t \right)(\Theta) = \left\{
\begin{array}{l@{\quad,\quad}c}
{\displaystyle \frac{d}{d \Theta}} \zeta_t(\Theta) & -1 \le \Theta < 0 \\[3mm]
{\cal L} [\zeta_t] & \Theta = 0
\end{array} \right. \, ,
\end{eqnarray}
where the linear functional is given by
\begin{eqnarray}
\label{g-wr1-syself}
{\cal L} \left[ \zeta_t \right] &=&
\int_{-1}^0 d \Theta \, \omega(\Theta) \zeta_t(\Theta)
\end{eqnarray}
with the scalar density
\begin{eqnarray}
\label{g-wr1-syseld}
\omega(\Theta) &=& \left.
\frac{\delta {\cal F}[z_t]}{\delta z_t(\Theta)}
\right|_{z_t(\Theta) = z_{\rm stat}^{\rm I} } = 
- R \, \mbox{\boldmath $\delta$}(\Theta + 1) \, .
\end{eqnarray}
Inserting the solution ansatz
\begin{eqnarray}
\zeta_t (\Theta) =  \phi^{\lambda} (\Theta) e^{\lambda t}\, , \quad -1 \le \Theta \le 0 
\end{eqnarray}
into (\ref{g-wr1-sysel}) leads to the eigenvalue problem of the infinitesimal generator ${\cal G}_L$:
\begin{eqnarray}
\label{EVP}
\lambda  \phi^{\lambda} (\Theta) = \left( {\cal G}_L \phi^{\lambda} \right) (\Theta)  \,, \quad -1 \le \Theta \le 0 \, .
\end{eqnarray}
Taking into account the definition of ${\cal G}_L$ in (\ref{g-wr1-sysel}), the eigenfunction $\phi^{\lambda} (\Theta)$
is determined to be 
\begin{eqnarray}
\label{uns}
\phi^{\lambda} (\Theta) = N_{\lambda} e^{\lambda \Theta} \,, \quad -1 \le \Theta \le 0 \,,
\end{eqnarray}
and the eigenvalue $\lambda$ follows from
\begin{eqnarray}
\lambda = L ( \lambda ) \, ,
\end{eqnarray}
where $L(\lambda)$ is defined by
\begin{eqnarray}
\label{lll}
L(\lambda)  = \int\limits_{-1}^0 d \Theta \, \omega (\Theta) e^{\lambda \Theta} \, .
\end{eqnarray}
Using the scalar density (\ref{g-wr1-syseld}), we obtain the following
transcendental characteristic equation:
\begin{eqnarray}
\label{g-wr1-char}
-R e^{-\lambda} - \lambda &=& 0 \, .
\end{eqnarray}
Thus the spectrum of the linear operator ${\cal G}_L$ has the following properties \cite{Hale}:
\begin{itemize}
\item It consists of a countable infinite number of eigenvalues
      which cumulate for $\Re(\lambda) \to -\infty$.
\item It is confined by an upper threshold for the real parts of
      the eigenvalues.
\item At the bifurcation point, i.e. the instability, some of the
      eigenvalues reach the imaginary axes and thus become unstable.
\end{itemize}
Further properties of the eigenvalues of the characteristic equation
(\ref{g-wr1-char}) follow from  the Hayes theorem which can be found in Ref. \cite{Hayes}. It states that
all solutions of the transcendental equation
\begin{eqnarray}
\label{g-Hayes}
p + q e^{-\lambda} - \lambda = 0
\end{eqnarray}
possess a negative real part if and only if (a) $p < 1$ and 
(b) $p \, < \, -q \, < \, \sqrt{a_1^2 + p^2}$.
Here $a_1$ represents the solution of the transcendental
equation $a_1 = p \tan (a_1)$ which lies in the interval $[0,\pi)$.
For the special case $p = 0$ one can show that $a_1$ is equal
to $\pi/2$. The shaded region in Fig.~\ref{b-hayes} represents that
region of the parameter space $p,q$ where both conditions of the Hayes theorem are fulfilled. The upper boundary line stems
from (a)  $p < 1$ and (b1) $p \, < \, -q$, whereas the lower boundary line follows from (a)  $p < 1$ and 
(b2) $-q \, < \, \sqrt{a_1^2 + p^2}$.\\

Comparing (\ref{g-wr1-char}) with (\ref{g-Hayes}) we obtain the identification $p = 0$ and $q = -R$.
Changing the control parameter $R$ from $0$ to $\pi/2$, the corresponding point in the
parameter space $p,q$ moves along the $q$-axis from
the point $q = 0$ to $q =-\pi/2$. At this critical
value it reaches the boundary of the shaded stability region, i.e. no longer all solutions of the characteristic equation
(\ref{g-wr1-char}) have a negative real part. Therefore an instability occurs at $R_c=\pi/2$.\\

\begin{figure}
\begin{center}
\setlength{\unitlength}{1mm}
\begin{picture}(120,80)
%\graphpaper[5](0,0)(120,80)
%\put(-5,58){\epsfig{figure=pll_ew.eps,width=75\unitlength,angle=-90}}
\put(0,84){\epsfig{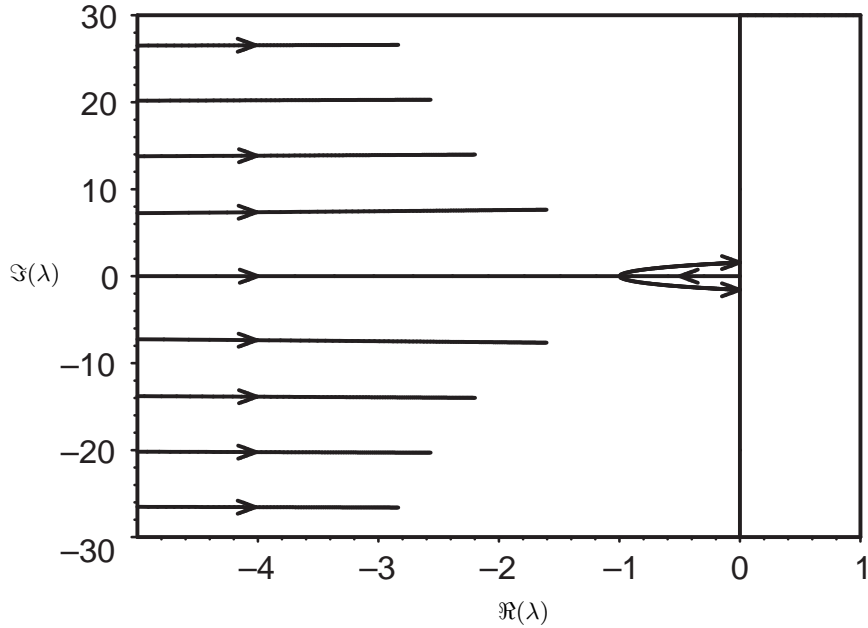}}
\put(60,-5){$\Re(\lambda)$}
\put(-5,40){$\Im(\lambda)$}
\end{picture}
$\mbox{}$\\[5mm]
\caption{\label{b-ew} The movement of the ten solutions of the characteristic equation (\ref{g-wr1-char}) with the
largest real part when the control parameter $R$ is increased from $0$ to $\pi/2$. At the instability $R_c=\pi/2$ two
complex conjugated eigenvalues have zero real part.}
\end{center}
\end{figure}

Fig.~\ref{b-ew} confirms this result by illustrating the movement of the ten solutions of the characteristic equation (\ref{g-wr1-char})
with the largest real part when the control parameter is increased from $0$ to $\pi/2$. The eigenvalues were obtained with a
Newton algorithm and the control parameter $R$ was increased in equidistant steps. For $R=0$ there exists only one real eigenvalue
$0$ as the linearized delay differential equation (\ref{g-wr1-sysel}) degenerates to an ordinary differential equation. For
$R>0$ this eigenvalue remains real and becomes negative. Furthermore, a countable infinite number of conjugate complex
eigenvalues and another real eigenvalue emerge from an infinite negative real part. At the value $R=1/\mbox{e}$ both real
eigenvalues meet at the point $(-1/0)$. They are converted to a pair of conjugated complex eigenvalues for 
$1/\mbox{e}\,<R<\pi/2$. These two complex conjugated
eigenvalues have zero real part at the instability $R_c=\pi/2$, which thus represents a Hopf bifurcation. 
We can further analyze this instability by introducing the smallness parameter
\begin{eqnarray}
\label{small}
\eps = \frac{R-R_c}{R_c} \quad \Longleftrightarrow \quad R = R_c ( 1 + \eps )
\end{eqnarray}
for the deviation from the critical control parameter $R_c=\pi/2$. In particular, we can determine both eigenvalues 
$\lambda_u^{\pm}(\eps)$ with nearly vanishing real part at $\eps \approx 0$ from the characteristic equation (\ref{g-wr1-char}):
\begin{eqnarray}
\label{eigen}
\lambda_u^{\pm}(\eps) =
\frac{R_c^2}{1+R_c^2} \, \eps \pm i R_c \left( 1 + \frac{1}{1+R_c^2} \,
\eps \right) + {\cal O} \left( \eps^2 \right) \, .
\end{eqnarray}
In the vicinity of the instability $\eps \approx 0$ we read off from Fig.~\ref{b-ew} that only the two eigenvalues (\ref{eigen})
have nearly vanishing real part, all other eigenvalues have a large negative real part:
\begin{eqnarray}
\Re[\lambda_u^{\pm} (\eps \approx 0)] \approx 0 \, ;\quad
\Re[\lambda_s^j(\eps \approx 0 )] < 0 \, , \quad j = 1 , \ldots , \infty \, .
\end{eqnarray}
This characteristic property of the linearized system (\ref{g-wr1-sysel}) leads to a the time-scale hierarchy
\begin{eqnarray}
\label{scale}
T_u^{\pm} = \frac{1}{\Re[\lambda_u^{\pm} (\eps \approx 0)]} \gg T_s^j = \frac{1}{\Re[\lambda_s^j(\eps \approx 0)]} \, , 
\quad j = 1 , \ldots , \infty \, .
\end{eqnarray}
Thus the infinite-dimensional extended state space ${\cal C}$ decomposes in an 2-dimensional subspace ${\cal U}$
of the linear unstable modes and a remaining infinite-dimensional subspace ${\cal S}$ of the linear stable modes
\cite{Wolfgang}. As a consequence, the extended state function $z_t (\Theta)$ 
can be decomposed near the instability according to 
\begin{eqnarray}
\label{g-wr1-zer1}
z_t(\Theta) = z_{\rm stat}^{\rm I}+ u_t(\Theta) + s_t(\Theta)=
u_t(\Theta) + s_t(\Theta) \, , \quad -1 \le \Theta \le 0 \, ,
\end{eqnarray}
as we have $z_{\rm stat}^{\rm I}=0$. Here $u_t(\Theta)$ and $s_t(\Theta)$ denote the respective contributions of $z_t ( \Theta)$ in
the subspaces ${\cal U}$ and ${\cal S}$. In order to project into these subspaces we need the linear unstable modes
\begin{eqnarray}
\label{g-wr1-ev}
\phi^{\lambda_u^{\pm}}(\Theta) =
N_{\lambda_u^{\pm}} e^{\lambda_u^{\pm} \Theta} \, , \quad -1 \le \Theta \le 0  
\end{eqnarray}
of the system (\ref{g-wr1-sysel}) which have been already determined in (\ref{uns}). 
However, their knowledge is not sufficient as the infinitesimal generator ${\cal G}_L$
is not selfadjoint. Therefore we need also the linear unstable modes
\begin{eqnarray}
\label{g-wr1-eva}
\psi^{\dagger \lambda_u^{\pm}}(s) =
N_{\lambda_u^{\pm}} e^{-\lambda_u^{\pm} s} \, ,  \quad 0 \le s \le 1 
\end{eqnarray}
of the adjoint system 
\begin{eqnarray}
\label{g-wr1-syselb}
\frac{d}{dt} \zeta_t^{\dagger} (s) &=&
-\left( {\cal G}_L^{\dagger} \, \zeta_t^{\dagger} \right)(s) = \left\{
\begin{array}{l@{\quad,\quad}c}
{\displaystyle \frac{d}{d s}} \zeta_t^{\dagger} (s) & 0 < s \le 1 \\[3mm]
-{\cal L}^{\dagger} [\zeta_t^{\dagger}] & s = 0
\end{array} \right. \, ,
\end{eqnarray}
where the linear functional
\begin{eqnarray}
\label{g-wr1-syselfb}
{\cal L}^{\dagger} \left[ \zeta_t^{\dagger} \right] &=&
\int_0^1 d s \, \omega(-s) \zeta_t^{\dagger} (s) 
\end{eqnarray}
contains also the scalar density (\ref{g-wr1-syseld}). Indeed, the solution ansatz
\begin{eqnarray}
\zeta_t^{\dagger} (s) = \psi^{\dagger \lambda} (s) e^{-\lambda s} \, , \quad 0 \le s \le 1 
\end{eqnarray}
converts (\ref{g-wr1-syselb}) to the eigenvalue problem
\begin{eqnarray}
\lambda  \psi^{\dagger} {}^{\lambda} (s) = \left( {\cal G}_L^{\dagger} \psi^{\dagger \lambda} \right) (s)  \, ,\quad 0 \le s \le 1 \, ,
\end{eqnarray}
which is adjoint to (\ref{EVP}). 
Note that $\psi^{\dagger \lambda}(s)$ and $\zeta_t^{\dagger} (s)$ are elements of the dual extended state space 
${\cal C}^{\dagger}$ which consists of continuous complex valued functions on the interval $[0,1]$. 
The relation between both extended state spaces ${\cal C}$ and ${\cal C}^{\dagger}$
is defined by the bilinear form \cite{Wolfgang}
\begin{eqnarray}
\label{form}
\left( \psi^{\dagger} | \phi \right) = \psi^{\dagger} ( 0 ) \phi ( 0 )  - \int\limits_{-1}^0 d \Theta
\int\limits_{0}^{\Theta} d s \, \psi^{\dagger} ( s - \Theta ) \omega ( \Theta ) \phi ( s )  \, .
\end{eqnarray}
Using this bilinear form one can show that the eigenfunctions
(\ref{g-wr1-ev}) and (\ref{g-wr1-eva}) are biorthonormal:
\begin{eqnarray}
\left( \psi^{\dagger}{}^{\lambda_i} | \phi {}^{\lambda_j} \right) = \delta_{ij} 
\, , \quad i , j = \pm \, .
\end{eqnarray} 
This determines the yet unknown normalization constants to be
\begin{eqnarray}
\label{g-wr1-norm}
N_{\lambda_u^{\pm}} &=& \frac{1}{\sqrt{1+\lambda_u^{\pm}}} \, ,
\end{eqnarray}
so they reduce near the instability because of (\ref{eigen}) to
\begin{eqnarray}
\label{norme}
N_{\lambda_u^{\pm}} &=&
\frac{1}{\sqrt{1 \pm i R_c}} + {\cal O} \left (\eps \right) \, .
\end{eqnarray}
Furthermore, the bilinear form (\ref{form}) allows to define the projector into the 2-dimensional subspace ${\cal U}$
of the unstable modes:
\begin{eqnarray}
\label{g-wr1-pu}
\left( {\cal P}_u \,\,\bullet \right) (\Theta)&=&
\sum_{i=\pm} \phi^{\lambda_u^i}(\Theta) \left( \left. \psi^{\dagger \lambda_u^i} \right| \bullet 
\right) \, .
\end{eqnarray}
Correspondingly, the projector into the remaining infinite-dimensional subspace ${\cal S}$ of the stable modes reads
\begin{eqnarray}
\label{g-wr1-ps}
{\cal P}_s \,\,\bullet &=&  \left( {\cal I} - {\cal P}_u \right) \bullet  \, .
\end{eqnarray}
Applying the projector ${\cal P}_u$ to $z_t \in {\cal C}$ leads to $u_t \in {\cal U}$ according to
\begin{eqnarray}
\label{propa}
u_t (\Theta) = \left( {\cal P}_u z_t \right) (\Theta) = \sum_{i=\pm} u^i ( t ) \phi^{\lambda_u^i} (\Theta) \, ,
\end{eqnarray}
where the amplitudes of the linear unstable modes $\phi^{\lambda_u^{\pm}} (\Theta)$ are defined by
\begin{eqnarray}
\label{propb}
u^{\pm} ( t ) = \left( \left. \psi^{\dagger \lambda_u^{\pm}} \right|  z_t \right)  \, .
\end{eqnarray}
Later on, these amplitudes represent the order parameters which indicate the emergence of an instability.
Analogously, the projector (\ref{g-wr1-ps}) leads to the stable modes
\begin{eqnarray}
\label{propc}
s_t (\Theta)  = \left( {\cal P}_s z_t \right) (\Theta) \,, \quad -1 \le \Theta \le 0 \, .
\end{eqnarray}
\section{Nonlinear Synergetic Analysis}

After having performed a linear stability
analysis around the reference state $z_{\rm stat}^{\rm I}=0$ in the vicinity of the instability
$R_c=\pi/2$, we now return to our original nonlinear evolution equation (\ref{g-wr1-syse}) in the extended state space ${\cal C}$.
We proceed by decomposing the generator ${\cal G}$ into its linear part ${\cal G}_L$ and a remaining effective nonlinear part:
\begin{eqnarray}
\label{g-wr1-sysee}
\frac{d}{dt} z_t(\Theta) &=&
\left( {\cal G}_L \, z_t \right)(\Theta)
+ X_0(\Theta) \cFeff [z_t]
\, , \quad -1 \le \Theta \le 0  \, .
\end{eqnarray}
Here we introduced the scalar function
\begin{eqnarray}
\label{scalar}
X_0(\Theta) &=& \left\{
\begin{array}{l@{\quad,\quad}c}
0 & -1 \le \Theta < 0 \\
1 & \Theta = 0
\end{array} \right.
\end{eqnarray}
and the effective nonlinear functional
\begin{eqnarray}
\label{effn}
\cFeff [z_t] &=&
\int_{-1}^0 d \Theta_1 \int_{-1}^0 d \Theta_2 \,
\omega^{(2)}(\Theta_1,\Theta_2) \,
z_t(\Theta_1) z_t(\Theta_2) 
\end{eqnarray}
with the scalar density (\ref{g-wr1-sysed2}). Using the projectors (\ref{g-wr1-pu}) and (\ref{g-wr1-ps}) and their properties
(\ref{propa})--(\ref{propc}), we can investigate the respective contributions of the order parameters $u^{\pm}(t)$
and the linear stable modes $s_t \in {\cal S}$
to the nonlinear dynamics (\ref{g-wr1-sysee}). Thus we obtain the following system of coupled nonlinear mode equations:
\begin{eqnarray}
\label{ueq}
\frac{d}{dt} u^{\pm} (t) & = & \lambda_u^{\pm} u^{\pm} ( t )+ \psi^{\dagger}{}^{\lambda_u^{\pm}} (0) \, \cFeff \left[ 
\sum_{j=\pm} \phi {}^{\lambda_u^j} u^j ( t ) + s_t \right] \, , \\
\label{seq}
\frac{d}{dt} s_t ( \Theta ) & = & \left( {\cal G}_L s_t \right) ( \Theta ) + \left( \left( {\cal I} -
{\cal P}_u \right) X_0 \right) (\Theta) \, \cFeff \left[ 
\sum_{j=\pm} \phi {}^{\lambda_u^j} u^j ( t ) + s_t \right] \, .
\end{eqnarray}
It is still exact and describes completely the nonlinear dynamics. However, a solution to these equations can be only found by
means of an approximation method. Such a well-established
approximative solution is provided by the slaving principle of synergetics
\cite{Haken1,Haken2,Haken3,Haken4,Haken5}. \\

To this end we start with the time-scale hierarchy (\ref{scale}) near the instability 
which leads to the fact that the dynamics of the
stable modes $s_t \in {\cal S}$ evolves much faster than the order parameters $u^{\pm}(t)$. In Ref. \cite{Wolfgang}
it has been shown for a quite general class of delay differential equations that such a time-scale hierarchy leads to a 
slaving of the stable modes, i.e. the numerous fast modes $s_t \in {\cal S}$ quasi-instantaneously take values which are
prescribed by the few slow order parameters $u^{\pm}(t)$. In our context, the slaving principle states mathematically that the
dynamics of the stables modes $s_t \in {\cal S}$ is determined by the center manifold $h(\Theta,u^+,u^-)$ according to
\begin{eqnarray}
\label{cenman}
s_t (\Theta) = h \left( \Theta, u^+(t),u^- ( t ) \right) \, .
\end{eqnarray}
Inserting this ansatz in (\ref{seq}) leads to an implicit equation for the center manifold $h(\Theta,u^+,u^-)$:
\begin{eqnarray}
&& \sum_{i=\pm} \frac{\partial h\left( \Theta, u^+ ( t ), u^- ( t ) \right)}{\partial u^i ( t )} \left(
\lambda_u^i u^i ( t ) + \psi^{\dagger}{}^{\lambda_u^i} (0) \, \,\cFeff \left[ 
\sum_{j=\pm} \phi {}^{\lambda_u^j} u^j ( t ) + h \right] \right) \nonumber \\
&& \hspace*{3cm} = \left( {\cal G}_L h\right) (\Theta) +
\left( \left( {\cal I} -
{\cal P}_u \right) X_0 \right) (\Theta) \, \cFeff \left[ 
\sum_{j=\pm} \phi {}^{\lambda_u^j} u^j ( t ) + h \right] \, .
\label{centere}
\end{eqnarray}
It can be approximately solved in the vicinity of the instability as follows. We assume that the order parameters $u^{\pm}(t)$
possess a certain dependence on the smallness parameter (\ref{small}) which is typical for a Hopf bifurctation:
\begin{eqnarray}
\label{epu}
u^{\pm} ( t ) = {\cal O} ( \eps^{1/2}) \, . 
\end{eqnarray}
Furthermore, we perform for the center manifold $h(\Theta,u^+,u^-)$ the lowest-order ansatz
\begin{eqnarray}
\label{g-wr1-zm1}
h(\Theta,u^+,u^-) =
\sum_{j_1=\pm} \sum_{j_2=\pm} H_{j_1 j_2}(\Theta) u^{j_1}(t) u^{j_2}(t) \, ,
\end{eqnarray}
as $r=2$ is the order of the effective nonlinear functional (\ref{effn}). From (\ref{epu}) and (\ref{g-wr1-zm1}) follows then in 
lowest order of $\eps$ that the effective nonlinear functional $\cFeff$ in (\ref{centere}) can be approximated by
\begin{eqnarray}
\cFeff \left[ \sum_{j=\pm} \phi {}^{\lambda_u^j} u^j ( t ) + h \right] \approx \sum_{j_1=\pm} \sum_{j_2=\pm}
F_{j_1 j_2}^{\rm eff} u^{j_1}(t)  u^{j_2}(t) \, , 
\end{eqnarray}
where the coefficients $F_{j_1 j_2}^{\rm eff}$ read
\begin{eqnarray}
\label{g-wr1-F}
\Feff_{j_1 j_2} =
\int_{-1}^0 d \Theta_1 \int_{-1}^0 d \Theta_2 \,
\omega^{(2)}\left( \Theta_1,\Theta_2 \right)
\phi^{\lambda_u^{j_1}}(\Theta_1) \phi^{\lambda_u^{j_2}}(\Theta_2) \, .
\end{eqnarray}
Taking into account (\ref{g-wr1-sysed2}) and (\ref{g-wr1-ev}), these coefficients turn out to be
\begin{eqnarray}
\label{g-wr1-F11}
\Feff_{++} = {\Feff_{--}}^* =
-R N_{\lambda_u^+}^2 e^{- \lambda_u^+} \, , \quad
\label{g-wr1-F12}
\Feff_{+-} = \Feff_{-+} =
-R N_{\lambda_u^+} N_{\lambda_u^-} e^{- \lambda_u^+} \, .
\end{eqnarray}
As a consequence, we conclude from (\ref{centere}) in lowest order of $\eps$ that the coefficients $H_{j_1 j_2}(\Theta)$
of the center manifold (\ref{g-wr1-zm1}) are given by
\begin{eqnarray}
\label{g-wr1-h}
H_{j_1 j_2}(\Theta) &=& \Feff_{j_1 j_2} K_{j_1 j_2}(\Theta) \, ,
\end{eqnarray}
where the coefficients $K_{j_1 j_2}(\Theta)$ follow from
\begin{eqnarray}
\label{hhkj}
K_{j_1 j_2}(\Theta) = \left( \left[ {\cal G}_L - \lam \right]^{-1} \left( {\cal P}_u X_0 - X_0 \right) \right) 
(\Theta) 
\end{eqnarray}
with the abbreviation
\begin{eqnarray}
\label{g-wr1-lam}
\lam = \sum_{k=1}^2 \lambda_u^{j_k} \, .
\end{eqnarray}
In Ref. \cite{Wolfgang} it is shown that the operator $\left[ {\cal G}_L - \lam \right]^{-1}$ has the explicit representation
\begin{eqnarray}
\left( \left[ {\cal G}_L - \lam \right]^{-1} \chi \right) (\Theta) = \int\limits_0^{\Theta} d s \, e^{\lam (\Theta - s)}
\chi (s) + [L(\lam)-\lam]^{-1} \left( \chi(0) - \int\limits_{-1}^0 d \Theta \int\limits_0^{\Theta} d s \,
e^{\lam (\Theta - s)} \omega (\Theta) \chi (s)\right) e^{\lam \Theta}  \, ,
\end{eqnarray}
where $L(\lambda)$ is already defined in (\ref{lll}). After some calculation, which also involves (\ref{g-wr1-syseld}), 
(\ref{g-wr1-ev}), (\ref{g-wr1-eva}), (\ref{g-wr1-pu}), and (\ref{scalar}), it thus follows that the coefficients 
(\ref{hhkj}) are given by
\begin{eqnarray}
\label{g-wr1-K}
K_{j_1 j_2}(\Theta) =
\sum_{j=\pm} \frac{{N_{\lambda_u^j}^2}}{\lambda_u^j - \lam}
e^{\lambda_u^j \Theta} - \frac{e^{\lam \Theta}}{L(\lam) - \lam} \, . 
\end{eqnarray}
Thus we obtain together with (\ref{g-wr1-syseld}), (\ref{lll}), and (\ref{g-wr1-lam}):
\begin{eqnarray}
\label{g-wr1-K11}
\hspace*{-10mm}
K_{++} (\Theta)= K_{--}^* (\Theta) &=&
-\frac{N_{\lambda_u^+}^2 e^{\lambda_u^+ \Theta}}{\lambda_u^+}
+\frac{N_{\lambda_u^-}^2 e^{\lambda_u^- \Theta}}{\lambda_u^- - 2 \lambda_u^+}
+\frac{e^{2 \lambda_u^+ \Theta}}
{R e^{-2 \lambda_u^+} + 2 \lambda_u^+} \, , \\
\label{g-wr1-K12}
\hspace*{-10mm}
K_{+-} (\Theta)= K_{-+} (\Theta) &=&
-\frac{N_{\lambda_u^+}^2 e^{\lambda_u^+ \Theta}}{\lambda_u^-}
-\frac{N_{\lambda_u^-}^2 e^{\lambda_u^- \Theta}}{\lambda_u^+}
+\frac{e^{(\lambda_u^+ + \lambda_u^-) \Theta}}
{R  e^{-(\lambda_u^+ + \lambda_u^-)} + \lambda_u^+ + \lambda_u^-} \, .
\end{eqnarray}
This completes the lowest-order result for the center manifold $h(\Theta,u^+,u^-)$ which is given by
(\ref{g-wr1-zm1}), (\ref{g-wr1-F11}), (\ref{g-wr1-h}), (\ref{g-wr1-K11}), and (\ref{g-wr1-K12}).\\ 

Thus we can now consider the order 
parameter equation (\ref{ueq}). In lowest order in $\eps$ we take into account (\ref{g-wr1-sysed2}), (\ref{g-wr1-syselb}), 
(\ref{effn}), (\ref{cenman}), and (\ref{g-wr1-zm1})
so that it reduces to
\begin{eqnarray}
\label{g-wr1-ordgl}
\hspace*{-10mm}
\frac{d}{dt} u^{\pm}(t) = \lambda_u^{\pm} u^{\pm}(t)
R N_{\lambda_u^{\pm}} \prod_{l=1}^2 \left[
\sum_{j=\pm} \phi^{\lambda_u^j}(\vartheta_l) u^j(t) + 
\sum_{j_1=\pm} \sum_{j_2=\pm} H_{j_1 j_2}(\vartheta_l) u^{j_1}(t) u^{j_2}(t) \right] \, ,
\end{eqnarray}
where we set
\begin{eqnarray}
\vartheta_l = \left\{ \begin{array}{l@{\quad,\quad}c}
                   -1 & l=1 \\
                   0  & l=2
                   \end{array} \right. \, .
\end{eqnarray}
Note that the order parameter equation (\ref{g-wr1-ordgl}) turns out to be an ordinary differential equation, i.e. it no longer
contains memory effects. Furthermore, we observe that the center manifold explicitly enters the order parameter equation
(\ref{g-wr1-ordgl}) as a direct consequence of the quadratic nonlinearity of the Wright equation (\ref{g-wr1-sys}).
We remark that this effect, which is essential for the present synergetic analysis, was neglected in the neurophysiological study 
in Ref.~\cite{Peter}. In the subsequent section 
we show how the order parameter equation (\ref{g-wr1-ordgl}) is converted to the 
normal form of a Hopf bifurcation.
\section{Normal Form}

Now we perform a nonlinear transformation of the order parameters which eliminates those terms which are irrelevant for the
normal form of a Hopf bifurcation. As far as the so-called {\it near identity transformation} 
and the theory of normal forms in general is concerned, we refer to the Refs.
\cite{Marsden,NF1,NF2}. The terms in (\ref{g-wr1-ordgl}) which are relevant for the normal form of a Hopf bifurcation read
\begin{eqnarray}
\label{g-wr1-normal1}
\hspace*{-10mm}
\frac{d}{dt} u^{\pm}(t) &=&
\lambda_u^{\pm} u^{\pm}(t) + q_0^{\pm} {u^{\pm}(t)}^2 +
q_1^{\pm} u^{\pm}(t) u^{\mp}(t) + q_2^{\pm} {u^{\mp}(t)}^2 +
k_1^{\pm} {u^{\pm}(t)}^2 u^{\mp}(t) \, ,
\end{eqnarray}
as we can neglect quartic terms and nonresonant cubic terms due to the rotating wave approximation. The 
respective coefficients in Eq. (\ref{g-wr1-normal1}) are given by
\begin{eqnarray}
q_0^{\pm} &=& - R N_{\lambda_u^{\pm}}
\phi^{\lambda_u^{\pm}}(-1) \phi^{\lambda_u^{\pm}}(0) =
- R {N_{\lambda_u^{\pm}}}^3 e^{- \lambda_u^{\pm}} \, , \label{KO1} \\
q_1^{\pm} &=& - R N_{\lambda_u^{\pm}} \left[
\phi^{\lambda_u^{\pm}}(-1) \phi^{\lambda_u^{\mp}}(0) +
\phi^{\lambda_u^{\pm}}(0) \phi^{\lambda_u^{\mp}}(-1) \right] = - R {N_{\lambda_u^{\pm}}}^2 N_{\lambda_u^{\mp}}
\left( e^{-\lambda_u^{\pm}} + e^{-\lambda_u^{\mp}} \right) \, , \\
q_2^{\pm} &=& - R N_{\lambda_u^{\pm}}
\phi^{\lambda_u^{\mp}}(-1) \phi^{\lambda_u^{\mp}}(0) =
-R N_{\lambda_u^{\pm}} {N_{\lambda_u^{\mp}}}^2 e^{- \lambda_u^{\mp}} \, , \\
k_1^{\pm} &=& - R N_{\lambda_u^{\pm}} \left\{
\phi^{\lambda_u^{\pm}}(-1) \left[ H_{+-}(0) + H_{-+}(0) \right] +
\phi^{\lambda_u^{\mp}}(-1) H_{++}(0) + \right. \nonumber \\
&& \left. \phi^{\lambda_u^{\pm}}(0) \left[ H_{+-}(-1) + H_{-+}(-1) \right] +
\phi^{\lambda_u^{\mp}}(0) H_{++}(-1) \right\} \, , \label{KO2}
\end{eqnarray}
where we did not write down the explicit form of $k_1^{\pm}$ for simplicity.
Then the previous order parameters $u^{\pm} ( t )$ are transformed to new order 
parameters $v^{\pm} ( t )$ by the near identity transformation
\begin{eqnarray}
\label{g-wr1-trafo}
u^{\pm}(t) &=& v^{\pm}(t) + \alpha_0^{\pm} {v^{\pm}(t)}^2 +
\alpha_1^{\pm} v^{\pm}(t) v^{\mp}(t) + \alpha_2^{\pm} {v^{\mp}(t)}^2 \, ,
\end{eqnarray}
with the yet-unknown coefficients $\alpha_0^{\pm}$, $\alpha_1^{\pm}$, and $\alpha_2^{\pm}$.
As the $u^{\pm} ( t )$ are small quantities in the vicinity of the instability, the same holds for the $v^{\pm} ( t )$.
Inserting (\ref{g-wr1-trafo}) in (\ref{g-wr1-normal1}), we obtain a system of ordinary differential equations of the form
\begin{eqnarray}
\label{meq}
M ( t ) \,  \frac{d}{dt} \left( \begin{array}{@{}c} v^{+} ( t ) \\ v^{-} ( t ) \end{array} \right)
=\left( \begin{array}{@{}c} w^{+} ( t ) \\ w^{-} ( t ) \end{array} \right) \, ,
\end{eqnarray}
where the matrix $M(t)$ is defined by
\begin{eqnarray}
M (t) &=& \left(
\begin{array}{cc}
1 + 2 \alpha_0^+ v^+(t) + \alpha_1^+ v^-(t) &
\alpha_1^+ v^+(t) + 2 \alpha_2^+ v^-(t) \\
\alpha_1^- v^-(t) + 2 \alpha_2^- v^+(t) &
1 + 2 \alpha_0^- v^-(t) + \alpha_1^- v^+(t)
\end{array}
\right) \, .
\end{eqnarray}
For simplicity, we don't write down the explicit form of $w^{+} ( t )$ and $w^{-} ( t )$, but we note that they contain
$v^{+} ( t )$ and $v^{-} ( t )$ at least in first order. Thus we obtain from (\ref{meq})
\begin{eqnarray}
\label{meqb}
\frac{d}{dt} \left( \begin{array}{@{}c} v^{+} ( t ) \\ v^{-} ( t ) \end{array} \right)
= M^{-1} (t)\left( \begin{array}{@{}c} w^{+} ( t ) \\ w^{-} ( t ) \end{array} \right) \, ,
\end{eqnarray}
with the inverse matrix
\begin{eqnarray}
M (t)^{-1} &=& \frac{1}{\mbox{Det}M (t)}
\left(
\begin{array}{cc}
M_{22}(t) & - M_{12}(t) \\
- M_{21}(t) & M_{11}(t)
\end{array}
\right) \, ,
\end{eqnarray}
where the determinant has the form
\begin{eqnarray}
\mbox{Det}\,\, M ( t ) = 1 + v^+(t) \left( 2 \alpha_0^+ + \alpha_1^- \right) +
2 v^+(t) v^-(t) \left( \alpha_0^+ \alpha_0^- - \alpha_2^+ \alpha_2^- \right) +
2 {v^+(t)}^2 \left( \alpha_0^+ \alpha_1^- - \alpha_1^+ \alpha_2^- \right) +
\, c.c. \, .
\end{eqnarray}
Expanding the right-hand side of (\ref{meqb}) in powers of $v^{+} ( t )$ and $v^{-} ( t )$ up to the third order, we yield
\begin{eqnarray}
\frac{d}{dt} v^{\pm}(t) &=&
\lambda^{\pm} v^{\pm}(t) +
\left( q_1^{\pm} - \alpha_0^{\pm} \lambda^{\pm} \right) {v^{\pm}(t)}^2 +
\left( q_0^{\pm} - \alpha_1^{\pm} \lambda^{\mp} \right) v^+(t) v^-(t) \, +
\Big[ q_2^{\pm} + \alpha_2^{\pm} ( \lambda^{\pm} - 2 \lambda^{\mp} ) \Big]
{v^{\mp}(t)}^2 
\nonumber \\ &&
+ \Big[ k_1^{\pm} + q_0^{\pm} ( \alpha_1^{\mp} - \alpha_0^{\pm} )
- q_0^{\mp} \alpha_1^{\pm} \, + 
q_1^{\pm} \alpha_1^{\pm} +
2 q_2^{\pm} \alpha_2^{\mp} - 2 q_2^{\mp} \alpha_2^{\pm} +
\alpha_1^{\pm} \alpha_1^{\mp} \lambda^{\pm} +
2 \alpha_2^{\pm} \alpha_2^{\mp} ( 2 \lambda^{\pm} - \lambda^{\mp} ) 
\nonumber \\ && 
+\alpha_0^{\pm} \alpha_1^{\pm}
( \lambda^{\pm} + 2 \lambda^{\mp} ) \Big] {v^{\pm}(t)}^2 v^{\mp}(t) \, .
\label{nf2e}
\end{eqnarray}
Now we can fix the yet-unknown coefficients $\alpha_0^{\pm}$, $\alpha_1^{\pm}$, and $\alpha_2^{\pm}$ of the 
near identity transformation (\ref{g-wr1-trafo}) in such a way that all quadratic terms vanish. This leads to the conditions
\begin{eqnarray}
\label{nf2f1}
\alpha_0^{\pm} = \frac{q_0^{\pm}}{\lambda_u^{\pm}} \, , \quad
\label{nf2f2}
\alpha_1^{\pm} = \frac{q_1^{\pm}}{\lambda_u^{\mp}} \, , \quad
\label{nf2f3}
\alpha_2^{\pm} = \frac{q_2^{\pm}}{2 \lambda_u^{\mp} - \lambda_u^{\pm}} \, .
\end{eqnarray}
Thus Eq. (\ref{nf2e}) reduces to the normal form of a Hopf bifurcation
\begin{eqnarray}
\label{g-wr1-normal2}
\frac{d}{dt} v^{\pm}(t) =
\lambda_u^{\pm} v^{\pm}(t) + b^{\pm} {v^{\pm}(t)}^2 v^{\mp}(t) \, ,
\end{eqnarray}
where the Hopf parameter $b^{\pm}$ is given by
\begin{eqnarray}
b^{\pm} = k_1^{\pm} + \frac{
q_0^{\pm} q_1^{\pm}(4 {\lambda_u^{\pm}}^2 - {\lambda_u^{\mp}}^2) +
q_1^{\pm} q_1^{\mp}(2 \lambda_u^{\pm} \lambda_u^{\mp} - {\lambda_u^{\mp}}^2) +
2 q_2^{\pm} q_2^{\mp} \lambda_u^{\pm} \lambda_u^{\mp}}
{\lambda_u^{\pm} \lambda_u^{\mp} ( 2 \lambda_u^{\pm} - \lambda_u^{\mp})} \, .
\end{eqnarray}
Taking into account (\ref{small}), (\ref{eigen}), (\ref{g-wr1-ev}), and (\ref{norme}) as well as 
(\ref{g-wr1-F11}), (\ref{g-wr1-h}), (\ref{g-wr1-K11}), and (\ref{g-wr1-K12}) together with 
(\ref{KO1})--(\ref{KO2}), this Hopf parameter $b^{\pm}$
reads in the vicinity of the instability:
\begin{eqnarray}
\label{HHO}
b^{\pm} = - \frac{R_c}{5 (1 + R_c^2)^{\frac{3}{2}}}
\left[ (3 R_c - 1) \pm i (R_c + 3) \right]
+ {\cal O} \left( \eps \right) \, .
\end{eqnarray}
Performing the ansatz
\begin{eqnarray}
\label{g-wr1-polar}
v^{\pm}(t) &=& r(t) e^{\pm i \varphi(t)} \, ,
\end{eqnarray}
the normal form (\ref{g-wr1-normal2}) is transformed to polar coordinates
\begin{eqnarray}
\frac{d}{dt} r(t) &=& r(t) \left[ \Re \left( \lambda_u^{\pm} \right)
+ \Re \left( b^{\pm} \right) r(t)^2 \right] \, , \\
\frac{d}{dt} \varphi(t) &=&
\label{g-wr1-osc-p1}
\pm \left[ \Im \left( \lambda_u^{\pm} \right)
+ \Im \left( b^{\pm} \right) r(t)^2 \right] \, .
\label{PHD}
\end{eqnarray}
Thus taking into account (\ref{eigen}) and (\ref{HHO}) near the instability, we result in the oscillatory solution
\begin{eqnarray}
\label{g-wr1-osc}
r_{\rm stat} &=& \sqrt{- \frac{\Re \left( \lambda_u^{\pm} \right)}
{\Re \left( b^{\pm} \right)} } =
\sqrt{\frac{5 R_c }{3 R_c - 1}} \, \sqrt[4]{1 + R_c^2} \, \sqrt{\eps} +
{\cal O}\left( \eps \right) \, , \\
\label{g-wr1-oscb}
\frac{d}{dt} \varphi(t) &=& \pm \left[ \Im \left( \lambda_u^{\pm} \right)
+ \Im \left( b^{\pm} \right) r_{\rm stat}^2 \right] 
= R_c - \frac{R_c}{3 R_c - 1} \, \eps + {\cal O}\left( \eps^2 \right) \, .
\end{eqnarray}
In order to compare this result with numerical simulations, we have to convert this oscillatory solution back to the 
original state space $\Gamma$. At first we observe that we obtain for $z_t \in {\cal C}$ 
from (\ref{g-wr1-zer1}), (\ref{propa}), (\ref{cenman}), and (\ref{g-wr1-zm1}) near the instability:
\begin{eqnarray}
z_t(\Theta) = \sum\limits_{j=\pm} \phi^{\lambda_u^j}(\Theta) u^j(t) +
\sum_{j_1=\pm} \sum_{j_2=\pm}  H_{j_1 j_2}(\Theta) u^{j_1}(t) u^{j_2}(t) \, .
\end{eqnarray}
Taking into account the near identity transformation (\ref{g-wr1-trafo}) together with (\ref{g-wr1-ev}), 
this yields up to the first order in $\eps$:
\begin{eqnarray}
\label{ttheta}
z_t(\Theta) =  N_{\lambda_u^+} e^{\lambda_u^+ \Theta} v^+(t)
+ N_{\lambda_u^-} e^{\lambda_u^- \Theta} v^-(t) + 
a_0(\Theta) {v^+(t)}^2 + a_1(\Theta) v^+(t) v^-(t)
+ a_2(\Theta) {v^-(t)}^2 \, ,
\end{eqnarray}
where the coefficients $a_0(\Theta)$, $a_1(\Theta)$, $a_2(\Theta)$ read 
\begin{eqnarray}
\label{g-wr1-A0}
a_0(\Theta) &=& \left[
N_{\lambda_u^+} e^{\lambda_u^+ \Theta} \alpha_0^+ +
N_{\lambda_u^-} e^{\lambda_u^- \Theta} \alpha_2^- + H_{++}(\Theta)
\right] \, , \\
\label{g-wr1-A1}
a_1(\Theta) &=& \left[
N_{\lambda_u^+} e^{\lambda_u^+ \Theta} \alpha_1^+ +
N_{\lambda_u^-} e^{\lambda_u^- \Theta} \alpha_1^- +
H_{+-}(\Theta) + H_{-+}(\Theta)
\right] \, , \\
\label{g-wr1-A2}
a_2(\Theta) &=& \left[
N_{\lambda_u^+} e^{\lambda_u^+ \Theta} \alpha_2^+ +
N_{\lambda_u^-} e^{\lambda_u^- \Theta} \alpha_0^- + H_{--}(\Theta)
\right] \, .
\end{eqnarray}
Due to the relation (\ref{map}) between $z(t) \in \Gamma$ and $z_t \in {\cal C}$, we conclude from (\ref{ttheta}):
\begin{eqnarray}
\label{g-wr1-z1}
z(t) = N_{\lambda_u^+} v^+(t) + N_{\lambda_u^-} v^-(t) + a_0(0) {v^+(t)}^2 + a_1(0) v^+(t) v^-(t)
+ a_2(0) {v^-(t)}^2 \, .
\end{eqnarray}
Near the instability we obtain from (\ref{norme}) 
\begin{eqnarray}
\label{NEX}
N_{\lambda_u^{\pm}} = \frac{1}{\sqrt[4]{1 + R_c^2}} e^{\pm i \psi_1} + {\cal O}(\eps) 
\end{eqnarray}
with some phase $\psi_1$, whereas (\ref{g-wr1-polar})--(\ref{PHD}) leads to 
\begin{eqnarray}
\label{vss}
v^{\pm}(t) = r_{\rm stat} e^{\pm i \varphi(t)} 
\end{eqnarray}
with the radius (\ref{g-wr1-osc}) and the phase
\begin{eqnarray}
\label{phase}
\varphi(t) = \Omega (\eps)t + \varphi_0  \,.
\end{eqnarray}
Here the frequency turns out to be
\begin{eqnarray}
\label{defo}
\Omega (\eps) =
R_c - \frac{R_c}{3 R_c - 1} \eps \, .
\end{eqnarray}
Furthermore, we yield from (\ref{g-wr1-A0})--(\ref{g-wr1-A2}) by taking into account
(\ref{g-wr1-F11}), (\ref{g-wr1-h}), (\ref{g-wr1-K11}), (\ref{g-wr1-K12}), and (\ref{NEX}) in the lowest order of $\eps$
\begin{eqnarray}
a_0(0) = \frac{1}{\sqrt{5(1+R_c^2)}}\, e^{i \psi_2} \, , \quad
a_1(0) = 0 \, , \quad
a_2(0) = \frac{1}{\sqrt{5(1+R_c^2)}} \,e^{-i \psi_2} \, ,
\end{eqnarray}
where $\psi_2$ denotes some phase.
Thus we obtain the following result for $z(t) \in \Gamma$ near the instability
\begin{eqnarray}
\label{g-wr1-z}
z(t) = c_0(\eps) + c_1(\eps) \cos \left[ \varphi(t) + \psi_1 \right] + c_2(\eps) \cos \left[ 2 \varphi(t) + \psi_2 \right] +
{\cal O} \left( \eps^{\frac{3}{2}} \right) \, ,
\end{eqnarray}
where the respective coeffcients read
\begin{eqnarray}
\label{defc}
c_0(\eps) = 0 \, , \quad
c_1(\eps) = 
2 \,\sqrt{\frac{5 R_c}{3 R_c - 1}} \sqrt{\eps} \, , \quad
c_2(\eps) = 2\, \frac{\sqrt{5} R_c}{3 R_c - 1} \eps \, .
\end{eqnarray}
Now we compare the oscillatory solution (\ref{g-wr1-vgl}), (\ref{g-wr1-vglb}), which was obtained from using the method of
averaging, with ours (\ref{phase}), (\ref{defo}), (\ref{g-wr1-z}), (\ref{defc}) by taking into account the critical value
(\ref{crit}) of the control parameter. We conclude that both results coincide in the lowest-order $\eps^{1/2}$, 
but our result is even correct up to the order $\eps$. \\

From the near identity transformation (\ref{g-wr1-trafo}) as well as from (\ref{g-wr1-osc}) and (\ref{vss}) we conclude
that the order parameters $u^{\pm}(t)$ turn out to be of the order $\eps^{1/2}$. This result is consistent with our original
assumption (\ref{epu}) which was the basis of our approximate solution of the implicit equation for the center manifold
(\ref{centere}) in the vicinity of the instability. Thus our synergetic system analysis is justified a posteriori by self
consistency.\\

Note that the same perturbative result (\ref{phase}), (\ref{defo}), 
(\ref{g-wr1-z}), (\ref{defc}) for the oscillatory solution above the Hopf bifurcation
can be derived with the multiple scaling method \cite{Michael1}. It represents a technical procedure to deduce the normal form,
once the bifurcation type is known, by using the knowledge how the respective quantities depend on the smallness parameter
$\eps= (R-R_c)/R_c$. Although the multiple scaling method has been originally developed for ordinary differential equations
\cite{ms1,ms2,ms3}, it can be also applied to delay differential equations (see, for instance, the treatment in Ref. \cite{elena}).
\begin{figure}
\setlength{\unitlength}{1mm}
\begin{center}
\epsfig{figure=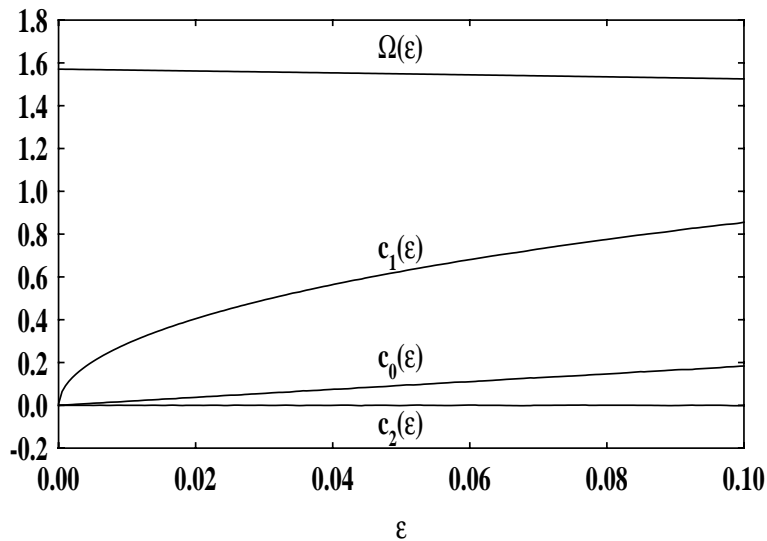,width=100\unitlength,angle=0}
\caption{\label{numana}Frequency $\Omega$ and Fourier coefficients $c_0$, $c_1$, $c_2$ of the oscillatory solution of the Wright
equation after the Hopf bifurcation versus the smallness parameter $\eps=(R-R_c)/R_c$. The interval of the smallness parameter
$[10^{-5},10^{-1}]$ was divided into 200 equidistant parts.}
\end{center}
\end{figure}
\section{Numerical Investigation}
\label{wr1-ver}

In order to numerically verify our analytical result, we integrated the underlying delay differential equation of Wright 
(\ref{g-wr1-sys}). By doing so, we varied the control parameter $R$ in the vicinity of the instability $R_c=\pi/2$ in such a way
that the smallness parameter $\eps = (R-R_c)/R_c$ took 200 equidistant values between $10^{-5}$ and $10^{-1}$. We used a
Runge-Kutta-Verner method of the IMSL library as an integration routine and performed a linear interpolation between the respective
values. In particular in the immediate vicinity of the instability the phenomenon of critical slowing down led to a 
transient behavior. To exclude this, we iterated the discretized
delay differential equation for each value of the control parameter at least
$10^6$ times. Afterwards we calculated the power spectrum with a complex FFT so that the basic frequency $\Omega$ of the oscillatory
solution could be determined with high resolution. Then we performed a real FFT with the period $T=2 \pi / \Omega$ of the simulated
periodic signal $z(t)=z(t+T)$:
\begin{eqnarray}
\label{F1}
z(t) = \frac{a_0}{2} + \sum_{k=1}^{\infty} \left[ a_k \cos \left( k \Omega t \right) + b_k \sin \left( k \Omega t \right) \right] \, .
\end{eqnarray}
The Fourier coefficients follow from integrations with respect to one period $T=2 \pi / \Omega$:
\begin{eqnarray}
a_k =  \frac{2}{T} \int\limits_0^{\infty} d t \, f ( t ) \,  \cos \left( k \Omega t \right) \, , \quad
k = 0 , 1 , \ldots , \infty \, ; \quad 
b_k  =  \frac{2}{T} \int\limits_0^{\infty} d t \, f ( t ) \,  \sin \left( k \Omega t \right) \, , \quad
k = 1 , \ldots , \infty \, . 
\end{eqnarray}
From (\ref{F1}) follows then the spectral representation
\begin{eqnarray}
\label{specrep}
z ( t ) = c_0 + \sum_{k=1}^{\infty} c_k \cos \left( k \Omega t + \phi_k \right) 
\end{eqnarray}
with the quantities
\begin{eqnarray}
c_0 =  \frac{a_0}{2} \, , \quad c_k = \sqrt{a_k^2 + b_k^2} \, , \quad \phi_k = - \mbox{artan}\,  \frac{b_k}{a_k} \, , 
\quad k = 1 , \ldots , \infty \, .
\end{eqnarray}
Thus our analytical result (\ref{phase}), (\ref{g-wr1-z}) 
can be interpreted as the first terms within a spectral representation (\ref{specrep}),
where the frequency $\Omega = 2 \pi / T$ and the Fourier coefficients $c_0$, $c_1$, $c_2$ are given by 
(\ref{defo}) and (\ref{defc}). Numerically
analysing the Hopf bifurcation with the FFT, the results for $\Omega$, $c_0$, $c_1$, $c_2$ are plotted in Fig.~\ref{numana} versus
the smallness parameter $\eps$. Comparing the respective numerical and analytical results,
we observe some deviations for small and for large values of the smallness parameter $\eps$. The former are
due to the phenomenon of critical slowing down, i.e. the system stays longer in the transient state when the instability 
is approached, and the latter arise from the neglected higher order corrections in the analytical approach. 
Therefore we restricted our numerical analysis to the
intermediate interval $[10^{-5},10^{-1}]$ of the smallness parameter $\eps$.
In Tab.~\ref{values} we see that the analytical and numerical determined quantities agree quantitatively very well. 
Thus our synergetic
system analysis for the delay-induced Hopf bifurcation in the Wright equation is numerically verified. \\

\begin{figure}
\setlength{\unitlength}{1mm}
\begin{center}
\vspace*{0.5cm}
\epsfig{figure=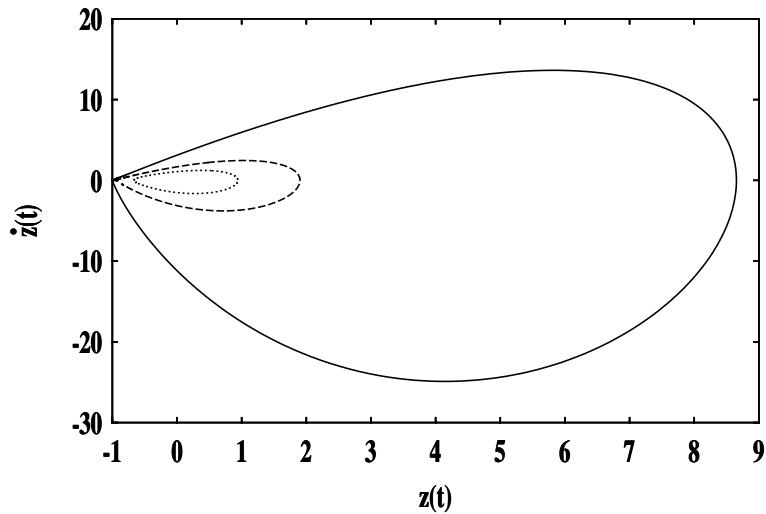,width=100\unitlength,angle=0}
\caption{\label{globif} Oscillatory solutions of the Wright equation (\ref{g-wr1-sys}) for three values of the control 
parameter $R$:
1.7 (dotted line), 2.0 (dashed line), and 3.247 (solid line).}
\end{center}
\end{figure}

For the sake of completeness we have also investigated oscillatory solutions for values of the
control parameter $R$ which are larger than the critical one $R_c=\pi/2$. Figure~\ref{globif} 
shows that all these periodic solutions oscillate around the stationary state $z_{\rm stat}^{\rm I}=0$ which becomes
unstable at $R_c=\pi/2$.
It turns out that a global bifurcation occurs for $R_c^g=3.247$ as then the
oscillatory solution comes close to the other stationary state  $z^{\rm II}_{\rm stat}=-1$ which turns out to be linear unstable for
all values of the control parameter $R>0$. 
Indeed, performing a linear stability analysis according to Section \ref{la} around the
stationary state $z^{\rm II}_{\rm stat}=-1$ leads to the characteristic equation
\begin{eqnarray}
\label{uuss}
R- \lambda =0 \, ,
\end{eqnarray}
so we have from Eq. (\ref{g-wr1-char}) the identification $p=R$ and $q=0$ 
(compare the shaded stability region in Fig.~\ref{b-hayes}).

\begin{table}
\begin{center}
\begin{tabular}{||c||c|c||c|c||c|c||}\hline
\begin{minipage}{2cm}
\begin{center}
investigated\\
quantity
\end{center}
\end{minipage} &
\multicolumn{2}{|c||}{
\begin{minipage}{2cm}
\begin{center}
analytical\\
expression
\end{center}
\end{minipage}} &
\multicolumn{2}{|c||}{
\begin{minipage}{3cm}
\begin{center}
analytical\\
value
\end{center}
\end{minipage}} &
\multicolumn{2}{|c||}{
\begin{minipage}{3cm}
\begin{center}
numerical\\
value
\end{center}
\end{minipage}} \\
& \hspace*{2mm}{\small axis intercept}\hspace*{2mm} & {\small slope}
& \hspace*{2mm}{\small axis intercept}\hspace*{2mm} & {\small slope} 
& \hspace*{2mm}{\small axis intercept}\hspace*{2mm} & {\small slope} \\ \hline \hline
\ru{-3}{0}{11}$\Omega(\eps)$ & $R_c$ & $  \hspace*{2mm} {\displaystyle \frac{R_c}{3 R_c -1}}$  \hspace*{2mm} &
$1.5708$ & \hspace*{2mm}$-0.4231$ \hspace*{2mm}& $1.5707$ & $-0.4024$ \\[4mm] \hline
\ru{-3}{0}{9}$c_0(\eps)$ & $0$ & $0$ & $0.0$ & $0.0$ &
$-2 \cdot 10^{-4}$ & $4 \cdot 10^{-2}$ \\[4mm] \hline
\ru{-3}{0}{11}$c_1(\eps)$ &
$ \hspace*{2mm}  2 \, {\displaystyle \sqrt{\frac{5 R_c}{3 R_c -1}}}$  \hspace*{2mm}  &  \hspace*{2mm} 
${\displaystyle \frac{1}{2}}$  \hspace*{2mm} &
$2.9090$ & $0.5$ & $2.890$ & \hspace*{3mm}$0.4999$\hspace*{3mm} \\[4mm] \hline
\ru{-3}{0}{11}$c_2(\eps)$ & $0$ & \hspace*{2mm} $2 \, {\displaystyle \frac{\sqrt{5} R_c}{3 R_c -1}}$  \hspace*{2mm} &
$0.0$ & $1.8923$ & $2 \cdot 10^{-4}$ & $1.832$ \\[4mm] \hline
\end{tabular}
\end{center}
\caption{\label{values} Comparison between the analytical and numerical values for the frequency $\Omega$ and the
Fourier coefficients $c_0$, $c_1$, $c_2$ of the oscillatory solution of the Wright equation after the Hopf bifurcation.}
\end{table}

\begin{figure}
\setlength{\unitlength}{1mm}
\begin{center}
\vspace*{-2cm}\epsfig{figure=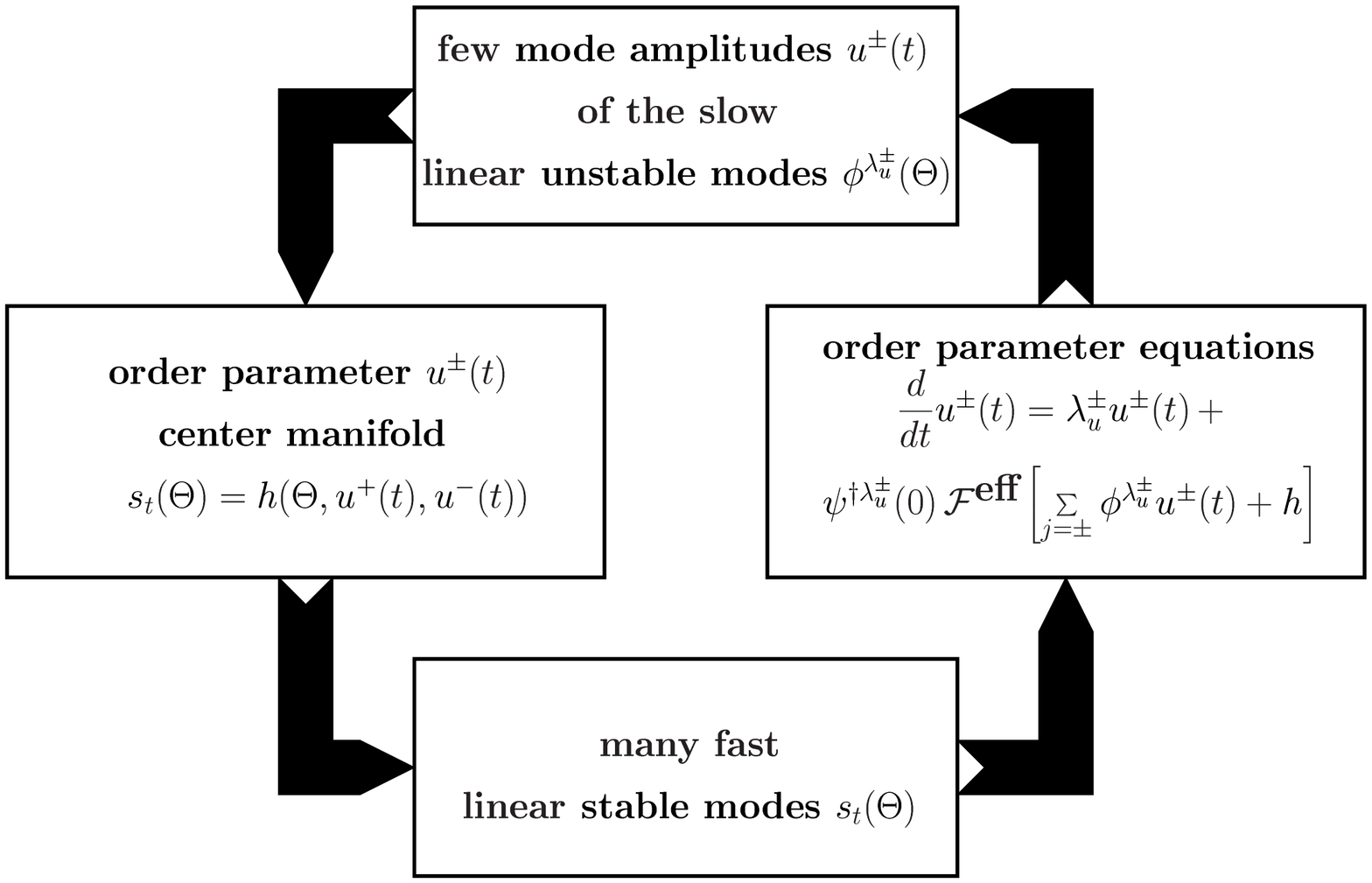,width=100\unitlength,angle=90}\vspace*{-1cm}
\caption{\label{circ} Circular causality chain of synergetics for the Hopf bifurcation of a delay differential equation.
On the one hand the center manifold of the slaving principle guarantees that many fast linear stable modes $s_t(\Theta)$
quasi-instantaneously take values which are prescribed by the few slow linear unstable modes $u^{\pm} (t)$. On the other hand
the adiabatic elimination of the fast enslaved modes $s_t(\Theta)$ influences the resulting order parameter equation.}
\end{center}
\end{figure}
\section{Summary and Outlook}

In this article a linear stability analysis of the Wright equation (\ref{g-wr1-sys}) around the stationary
state $z^{\rm I}_{\rm stat}=0$ showed that a delay-induced Hopf bifurcation occurs at the critical value $R_c=\pi/2$ 
of the control parameter $R$. Within a subsequent nonlinear synergetic analysis
we adiabatically eliminated the stable modes and derived the normal form of this Hopf bifurcation. It is explicitly influenced
by the center manifold in the lowest order,
as the Wright equation (\ref{g-wr1-sys}) has a quadratic nonlinearity. Solving the normal form
we obtained a periodic solution above the Hopf bifurcation which was numerically verified.\\

In contrast to the
corresponding analysis of the electronic system of a first-order phase-locked loop 
with time delay \cite{Michael2}, this paper confirms not only
the order parameter concept for delay systems. It also represents a successful test for the 
slaving principle of synergetics, i.e. for
the influence of the center manifold on the order parameter equations. Thus the validity of the circular causality chain
of synergetics (see Fig.~\ref{circ})
has been demonstrated for the Hopf bifurcation of a delay differential equation.\\

It remains to investigate the circular causality chain also for other bifurcations. The Floquet theory for delay
differential equations and thus the linear stability analysis for a periodic reference state was already established in
the Refs.~\cite{Christian1,Christian2}. However, a corresponding synergetic system analysis is still missing which derives
the order parameter equations and the normal forms for bifurcations of oscillatory solutions \cite{Michael3}.
\section*{Acknowledgements}
We are thankful to Michael Bestehorn and Rudolf Friedrich
for contributing various useful comments at an initial stage of this
work. Furthermore we thank Hermann Haken and Arne Wunderlin for teaching us synergetics for many years.
Finally, A.P. is grateful for the hospitality of G\"unter Wunner at the I. Institute for Theoretical Physics
of the University of Stuttgart as this article was finished there.
\end{document}